\documentclass[twocolumn]{aastex61}
\usepackage{amsmath,amstext}
\usepackage[T1]{fontenc}
\usepackage{apjfonts} 
\usepackage[figure,figure*]{hypcap}
\usepackage{breakurl}
\usepackage{color}

\newcommand{\arcsecs}{\hbox{$^{\prime\prime}$}}

\shorttitle{Automated Swirl Detection Algorithm (ASDA)}
\shortauthors{Jiajia Liu et al.}

\begin{document}

\title{Automated Swirl Detection Algorithm (ASDA) and Its Application to Simulation and Observational Data}
\author{Jiajia Liu}
\affiliation{Solar Physics and Space Plasma Research Center (SP2RC), School of Mathematics and Statistics, The University of Sheffield, Sheffield S3 7RH, UK}
\email{jj.liu@sheffield.ac.uk}

\author{Chris J. Nelson}
\affiliation{Solar Physics and Space Plasma Research Center (SP2RC), School of Mathematics and Statistics, The University of Sheffield, Sheffield S3 7RH, UK}
\affiliation{Astrophysics Research Centre (ARC), School of Mathematics and Physics, Queen's University, Belfast, BT7 1NN, Northern Ireland, UK.} 
\author{Robertus Erd\'elyi}
\affiliation{Solar Physics and Space Plasma Research Center (SP2RC), School of Mathematics and Statistics, The University of Sheffield, Sheffield S3 7RH, UK}
\affiliation{Department of Astronomy, E\"{o}tv\"{o}s Lor\'{a}nd University, Budapest, P\'{a}zm\'{a}ny P. s\'{e}t\'{a}ny 1/A, H-1117, Hungary}

\begin{abstract}
Swirling motions in the solar atmosphere have been widely observed in recent years and suggested to play a key role in channeling energy from the photosphere into the corona. Here, we present a newly-developed Automated Swirl Detection Algorithm (ASDA) and discuss its applications. ASDA is found to be very proficient at detecting swirls in a variety of synthetic data with various levels of noise, implying our subsequent scientific results are astute. Applying ASDA to photospheric observations with a spatial resolution of 39.2 km sampled by the Solar Optical Telescope (SOT) on-board Hinode, suggests a total number of $1.62\times10^5$ swirls in the photosphere, with an average radius and rotating speed of $\sim290$ km and $< 1.0$ km s$^{-1}$, respectively. Comparisons between swirls detected in Bifrost numerical MHD simulations and both ground-based and space-borne observations, suggest that: 1) the spatial resolution of data plays a vital role in the total number and radii of swirls detected; and 2) noise introduced by seeing effects could decrease the detection rate of swirls, but has no significant influences in determining their inferred properties. All results have shown that there is no significant difference in the analysed properties between counter-clockwise or clockwise rotating swirls. About 70\% of swirls are located in intergranular lanes. Most of the swirls have lifetimes less than twice of the cadences, meaning future research should aim to use data with much higher cadences than 6 s. In the conclusions, we propose some promising future research applications where ASDA may provide useful insights.
\end{abstract}


\section{Introduction} \label{intro}

Rotational motions have been observed on the Sun on a variety of scales, including sub-arcsecond vortex flows \citep{Bonet08}, large-scale sunspot rotations \citep{Evershed10,Brown03}, {tornadoes \citep{Li12, Wedemeyer12, Su12, Panesar2013, Wang16}}, spicules/jets \citep{Kitiashvili13, LiuW09, LiuJ14} and even coronal mass ejections (CMEs) \citep{Vourlidas11}. The small-scale photospheric vortices are widely hypothesised to form as a natural consequence to granular flows \citep{Wang95, Attie09} in the quiet Sun and could play a key role in the supply of energy to the upper solar atmosphere either, for example, through the build up of magnetic energy in twisted field lines in the corona or through the channeling of magneto-hydrodynamic (MHD) waves \citep{Velli99, Shelyag11, Shelyag13}. The upward propagation of wave energy along magnetic field lines due to torsional motions in the photosphere has been studied in a number of simulations \citep[e.g.,][]{Mumford15a, Mumford15b, Murawski18}.

The identification of swirls in the chromosphere was achieved by \cite{Wedemeyer09}, who observed swirls in chromospheric Ca II $8542$ \AA\ filtergrams. This discovery was helped greatly by the development of high-resolution Fabry-Perot instruments such as the {\it CRisp Imaging SpectroPolarimeter} (CRISP) \citep{Scharmer06,Scharmer08} at the Swedish Solar Telescope (SST) \citep{Scharmer03}. It was suggested that the observed swirls were evidence of twisted or twisting magnetic field lines in the upper atmosphere, with the twist being induced by convective buffeting in the photosphere \citep{Wedemeyer09}. A classification of multiple types of swirls was conducted by \cite{Wedemeyer13}.

Confirmation that chromospheric swirls could be a viable energy supply mechanism to the upper solar atmosphere was shown by \cite{Wedemeyer12}, who identified signatures of $14$ chromospheric swirls in the transition region and corona, as observed by the {\it Atmospheric Imaging Assembly} (AIA) \citep{Lemen12}, and named the entire structure (from the photosphere to the corona) `solar tornadoes'. The signatures of small-scale swirls in data sampled by the Interface Region Imaging Spectrograph (IRIS) \citep{dePontieu14} were then presented by \cite{Park16}, who found up-flow velocities of around $8$ km s$^{-1}$ during the course of the swirl but, interestingly, no signature in transition region spectral lines. It is unclear whether the event studied is a solar tornado or merely a chromospheric swirl.

Most of the previous reported swirls were detected manually, meaning human biases and limitations could result in underestimating the total number of swirls. Recently, attempts have been made to detect swirls using automated methods. \cite{Kato17} tested two automated detection routines on a simulated chromospheric time-series produced by the CO$^5$BOLD \citep{Freytag12} code. \cite{Kato17} found, using their preferred method (the vorticity strength method), average lifetimes of $52$ s and radii of $338$ km for swirls and suggested, therefore, that only a small fraction of the total number of swirl events that occur in the solar atmosphere might be observed using modern instrumentation. Specifically, extremely high cadence data would be required to identify any swirls corresponding to the simulated events. {On the other hand, although the vorticity strength method in \cite{Kato17} was found effective and accurate in determining the locations of swirls with almost 100\% detection rate and $>95\%$ location accuracy, the method cannot give any information of the edge of swirls, which can be seen from the relatively poor diameter accuracy (60\%-80\%) in Table 1 therein. Improved methods are needed to perform automated detection of swirls with both centers and edges accurately provided simultaneously. This aspect may be important if swirl detection is studied at multi-layers of the solar atmosphere, in particular for obtaining overlap statistics or swirl propagation. }

In this article, we report the release of an open-source Automated Swirl Dectection Algorithm (ASDA, available via \url{https://github.com/PyDL/ASDA}) written in the Python language, and its application to simulation and observational data. We set out our work as follows: In Sect.~\ref{met}, we outline the algorithm employed by ASDA and the validation of the code using a variety of synthetic data; Sect.~\ref{app} presents scientific results of the vortices detected based on both numerical simulation and observational data; and, Sect.~\ref{conc} contains our discussions and conclusions.

\begin{figure*}[htb]
\centering
\includegraphics[width=0.8\hsize]{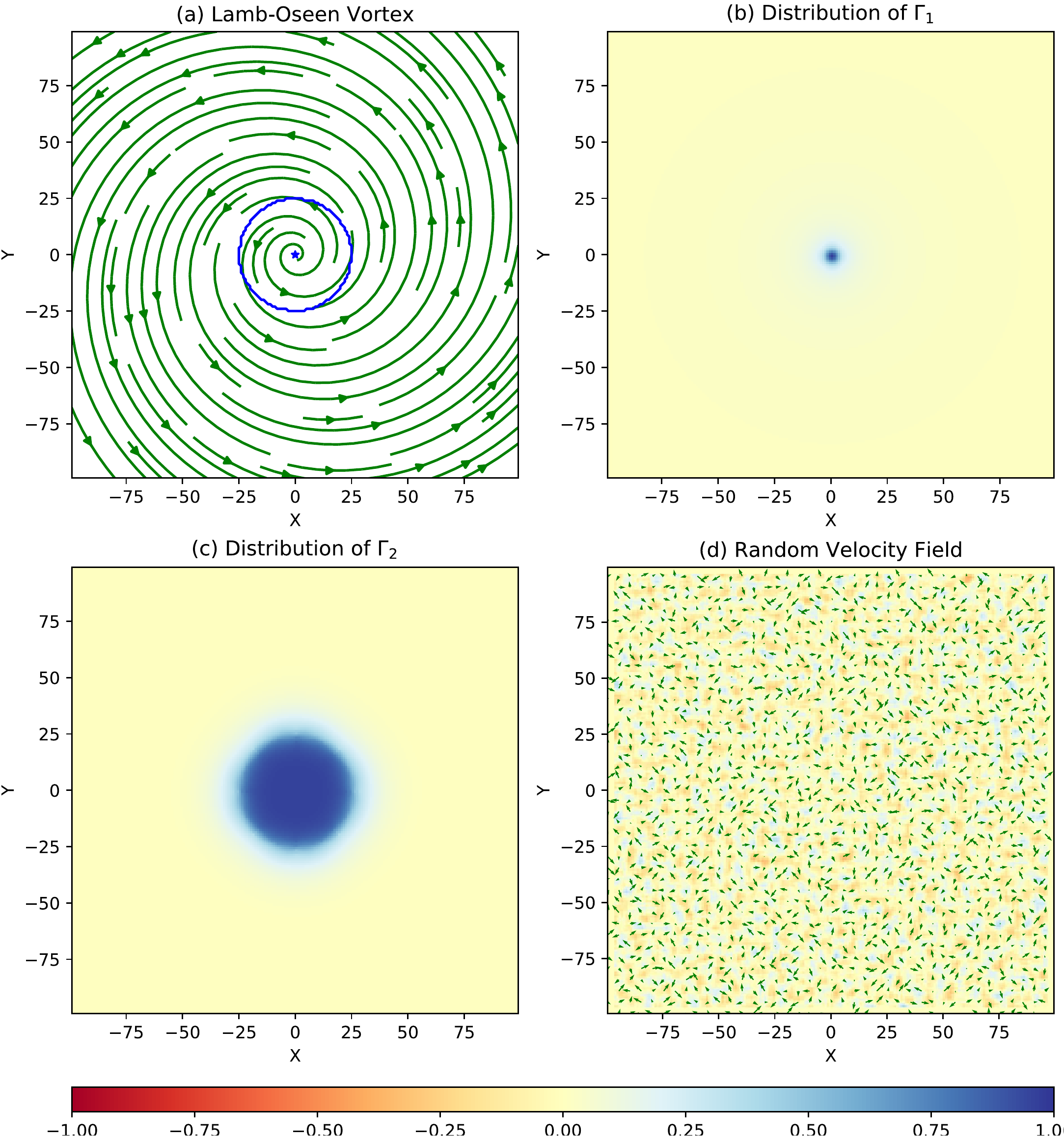}
\caption{\textbf{(a)} Velocity field of the generated Lamb-Oseen vortex (green arrows) and the detected vortex edge (blue circle) and center (blue asterisk) using ASDA. \textbf{(b)} and \textbf{(c)} Distributions of $\Gamma_1$ and $\Gamma_2$ of the generated Lamb-Oseen vortex in (a). \textbf{(d)} Velocity field of a random white noise velocity field (green arrows) and its corresponding $\Gamma_1$ distribution (yellow-white background). As expected, no swirls are detected in this random flow field}
\label{fig:test}
\end{figure*}

\section{Method} \label{met}
In brief, the basic workflow of ASDA contains two essential steps which are both required to perform swirl detections on preprocessed, scientifically-ready dataset from observations or simulations. These steps are: 1) the estimation of velocity field using the Fourier Local Correlation Tracking (FLCT) method \citep{Welsch04, Fisher08}; and 2) the application of vortex identification algorithms proposed by \cite{Graftieaux01} to the velocity field estimated in the first step. 

In this section, we describe the steps employed by ASDA in detail and test its reliability on a series of synthetic datasets.

\subsection{Swirl Detection Algorithms} \label{alg}
Current observational constrains mean it is currently not possible to directly detect the horizontal velocity field in the solar atmosphere. As such, various optical flow techniques have been developed in order to estimate the bulk velocity field from observables including intensities and magnetic fields. One widely used and accepted velocity-estimation techniques is known as the Local Correlation Tracking (LCT) method, which was firstly proposed by \cite{November88} and used to measure motions of solar granulation. This approach could be applied to either intensity or line-of-sight magnetic field observations \citep[e.g.][]{Fan12}.

As a part of the open-source ASDA code, we have also developed an integrated Python wrapper for the FLCT code. The Python code and its descriptions are available at \url{https://github.com/PyDL/pyflct}. The default size of the Gaussian apodizing window (the region considered as the ``local neighborhood'') is set to 10 pixels in the code, following the suggestion by \cite{Louis15}.

\begin{figure*}[htb]
\centering
\includegraphics[width=0.8\hsize]{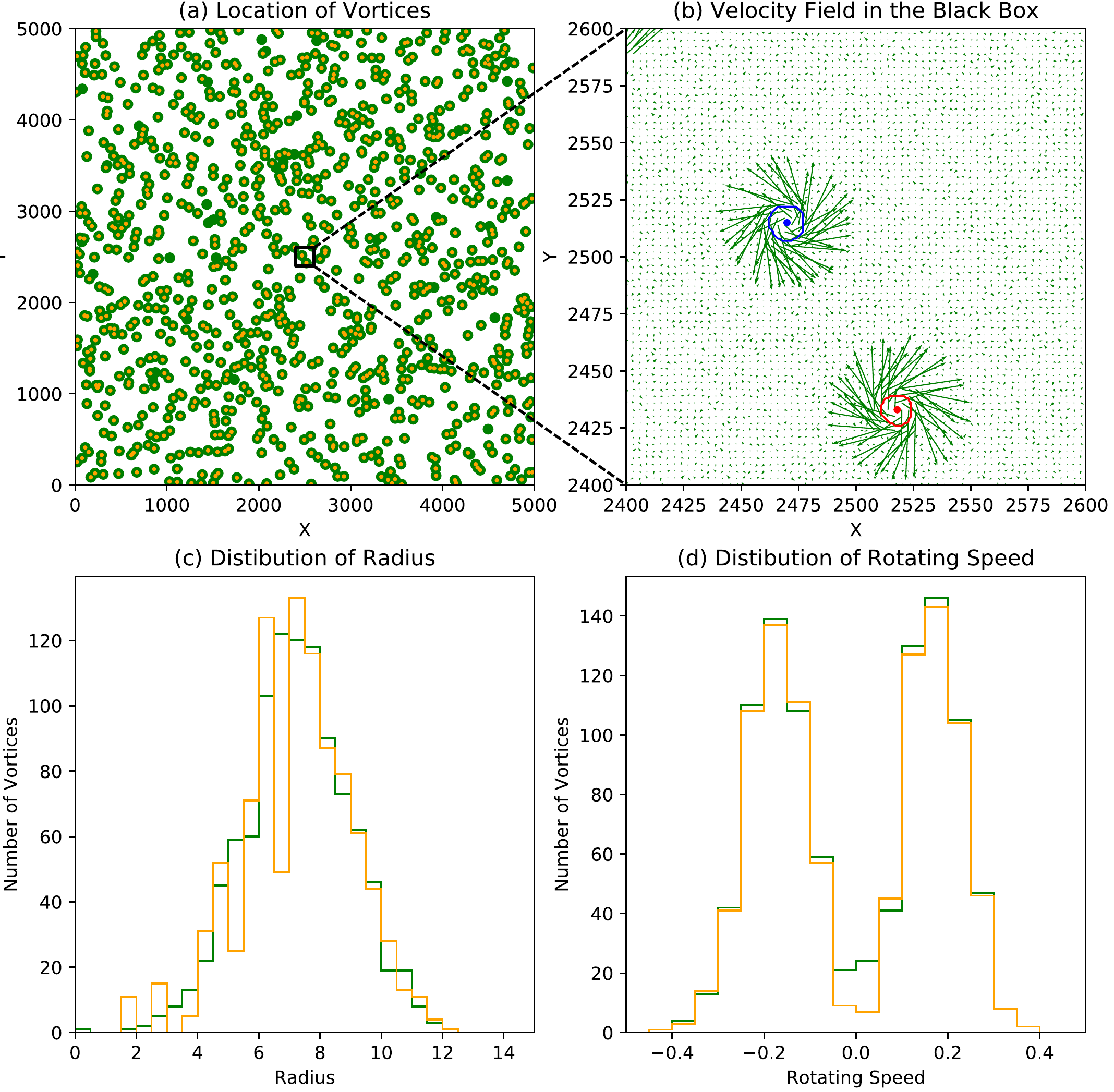}
\caption{\textbf{(a)} Locations of generated vortices (green dots) and detected vortices (orange dots) in the synthetic dataset with a noise level of 10\%. \textbf{(b)} Close-up of the black box in panel (a). Green arrows are velocity fields. Blue (red) dots and circles are the centers and edges of detected vortices with anti-clockwise (clockwise) rotation, respectively. \textbf{(c)} and \textbf{(d)} Histogram of the radius (c) and rotating speed (d) of the generated (green) and detected (orange) vortices.}
\label{fig:syn}
\end{figure*}

After obtaining the velocity field, we employ the vortex identification algorithms proposed by \cite{Graftieaux01} to identify swirls. For each pixel $P$,  two dimensionless parameters are defined as:

\begin{equation}
\begin{aligned}
\Gamma_1(P) = \frac{1}{N} \sum_S{\frac{\mathbf{n}_{PM} \times \mathbf{v}_M}{|\mathbf{v}_M|}}, \\
\Gamma_2(P) = \frac{1}{N} \sum_S{\frac{\mathbf{n}_{PM} \times (\mathbf{v}_M - \mathbf{\overline{v}})}{|\mathbf{v}_M  - \mathbf{\overline{v}}|}}.
\end{aligned}
\label{eq:gamma}
\end{equation}

\noindent Here, $S$ is a two-dimensional region with size $N$ pixels surrounding the target point $P$. $M$ is a point within the region $S$. $\mathbf{n}_{PM}$ is the normal vector pointing from point $P$ to $M$.  $\mathbf{v}_M$ is the velocity vector at point $M$. $\mathbf{\overline{v}}$ is the average velocity within the region $S$. $\times$ and $|\ |$ denote the cross product and the mode of vectors, respectively.

It is demonstrated in \cite{Graftieaux01} that, $|\Gamma_1|$ reaches its maximum value at the center of a vortex (Fig.~\ref{fig:test}b). Typically, the maximum value of $|\Gamma_1|$ at the center of a vortex should range from 0.9 to 1.0, where positive (negative) $\Gamma_1$ indicates a counter-clockwise (clockwise) rotating direction of the vortex. Practically, in our code, to allow some numerical errors, any point with $|\Gamma_1|$ less than 0.89 will not be considered as the center of a vortex. This means that any vortex-like structure with expanding/shrinking speed higher than half of its rotating speed will not be considered as a swirl by ASDA. Considering that the velocity field is locally dominated by rotation inside and by strain outside a vortex, \cite{Graftieaux01} found $|\Gamma_2|$ reaches a value of $2/\pi$ at the edge of the vortex, where again positive (negative) $\Gamma_2$ indicates a counter-clockwise (clockwise) rotating direction of the vortex.

\begin{table*}[ht]
\centering
\begin{tabular}{|c|c|c|c|c|c|}
\hline
Noise Level & Detection Rate & False Detection Rate & Location Accuracy  & Radius Accuracy & Rotating Speed Accuracy \\
\% & \% & \% & \% & \% & \% \\
\hline
0 & 99.8 & 0.0 & 100.0 & 96.5 & 99.4 \\
10 & 96.3 & 0.0 & 100.0 & 96.7 & 98.8 \\
25 & 85.5 & 0.0 & 99.9 & 93.6 & 98.0 \\
50 & 43.1 & 0.0 & 99.8 & 87.0 & 96.6 \\
75 & 14.0 & 0.0 & 99.6 & 79.8 & 93.6 \\
100 & 3.2 & 0.0 & 99.6 & 75.5 & 91.4 \\
\hline
\end{tabular}
\caption{\label{tab:syn} Average detection rate, false detection rate, location accuracy, radius accuracy and rotating speed accuracy of the detection on all 1000 inserted vortices in the synthetic data, with velocity noise levels of the background ranging from 0 to 100\%.}
\end{table*}

\begin{figure*}[htb]
\centering
\includegraphics[width=\hsize]{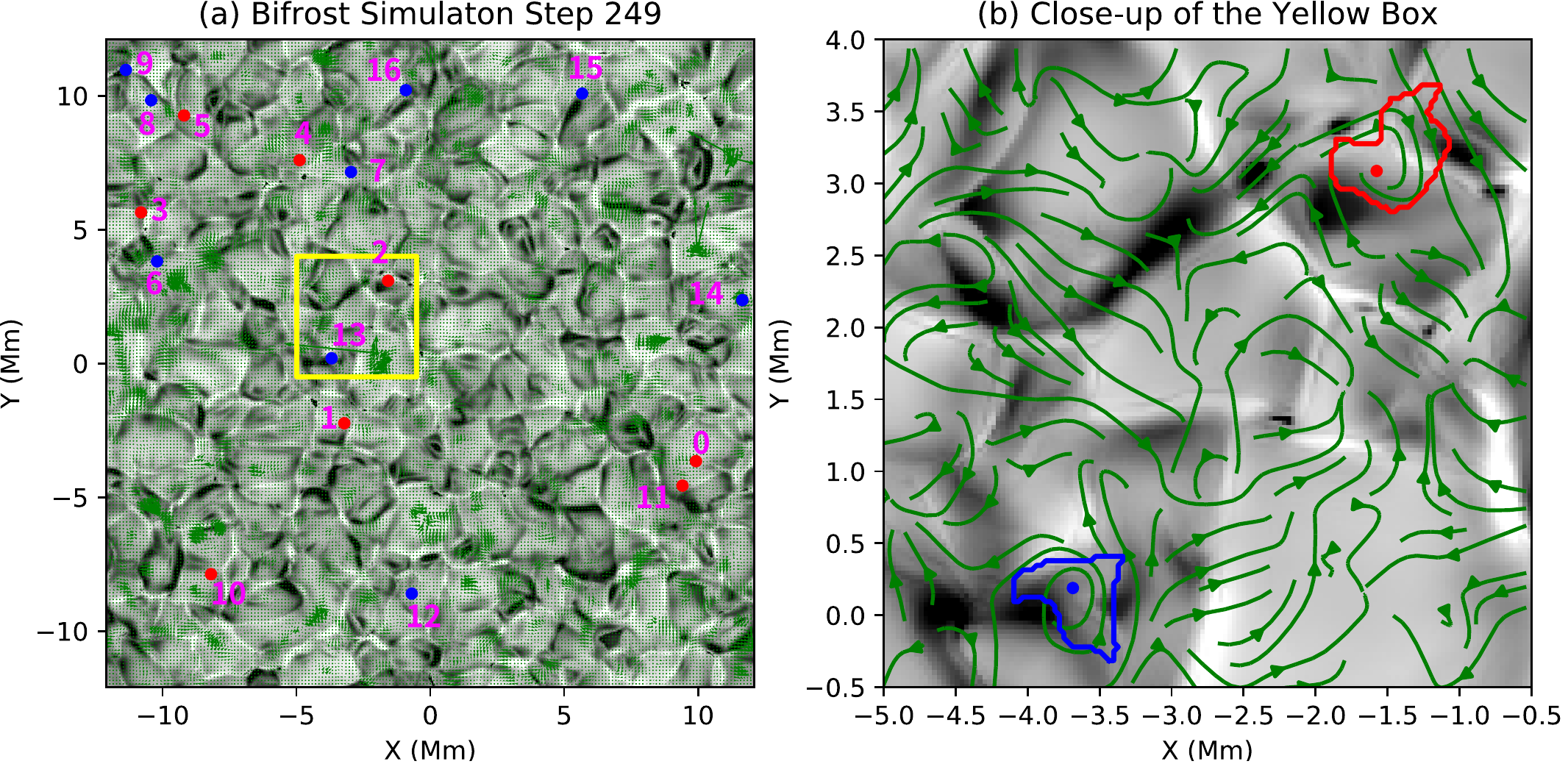}
\caption{\textbf{(a)} Mass density distribution in the photosphere (black-white) at time step 249 of the Bifrost simulation, the velocity field estimated by FLCT (green arrows) and the locations of the centers of detected photospheric swirls (blue and red dots) are overlaid. Numbers in purple are the sequential numbers of the detected swirls. \textbf{(b)} Close-up view of the yellow box in panel (a). Blue and red curves indicate the edges of the detected swirls.}
\label{fig:bifrost_full}
\end{figure*}

$N$, the number of points within the region $S$ in Eq.~\ref{eq:gamma}, has limited influence on the peak values of $\Gamma_1$ and $\Gamma_2$ \citep{Graftieaux01}. However, it can serve as a spatial filter. Considering that,  in observations, we need at least two pixels to confirm a reliable spatial structure and  at least two pixels to perform the FLTC velocity field estimation, we specify $N$ as 49 in our code. The above specification results in a $7\times7$ px$^2$ square region of $S$ and weakens swirls with radius less than 4 px. Moreover, any swirl detected in Sect.~\ref{app} with radius less than 4 px will be removed from the database due to the same reason. The Python code of the vortex detection algorithms can be found at \url{https://github.com/PyDL/ASDA}.

\subsection{Validation with Synthetic Data} \label{val}

\begin{figure*}[htb]
\centering
\includegraphics[width=0.8\hsize]{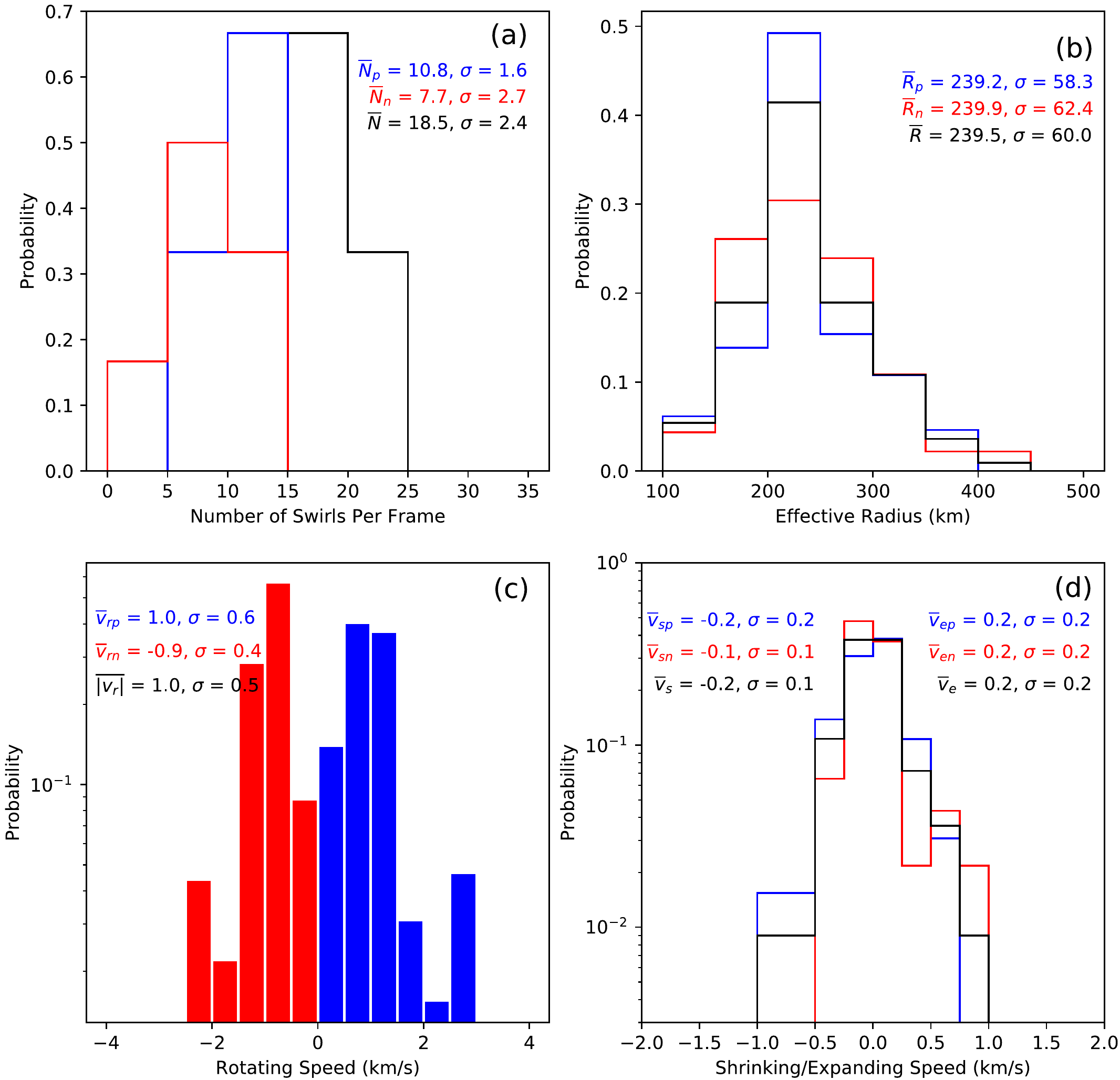}
\caption{Statistics of the number per frame ($N$), effective radius ($R$, Eq.~~\ref{eq:er}), rotating speed ($v_r$) and shrinking/expanding ($v_s$/$v_e$) speed of all 111 photospheric swirls detected by ASDA from the full-resolution Bifrost simulation data. Subscripts $p$ and $n$ (blue and red) denote positive and negative swirls, repectively. }
\label{fig:bifrost_stat_full}
\end{figure*}

Before applying ASDA to realistic numerical simulation and observational data, we first validate its reliability and examine its performance using synthetic data. In the first test we applied,  ASDA was used to detect a single positive (anti-clockwise) Lamb-Oseen vortex \citep{Saffman93} with maximum rotating speed $v_{max} = 25$ and radius $r_{max} = 5$. By definition, a point with a distance $r$ away from the center of a Lamb-Oseen vortex has a rotating speed:

\begin{equation}
v_r = v_{max}(1+\frac{1}{2\alpha})\frac{r_{max}}{r}[1-exp(-\alpha\frac{r^2}{r_{max}^2})],
\label{eq:lamb}
\end{equation}

\noindent where, $\alpha \approx 1.256$. Furthermore, we define the expanding speed $v_e$ at any point as 20\% of its rotating speed:

\begin{equation}
v_e = 0.2v_r,
\label{eq:lamb2}
\end{equation}

\noindent leading to an expanding speed of 1.0 at the edge of the vortex. The blue asterisk and circle in Fig.~\ref{fig:test} represent the detected center and edge of the above Lamb-Oseen vortex. The detected center locates exactly at the center of the vortex. The estimated radius, rotating and expanding speed of the vortex detected by ASDA are 25.2, 5.0, and 1.0 respectively, which are consistent with the defined radius ($r_{max}$), rotating speed ($v_{max}$) and expanding speed ($v_e$). Panels (b) and (c) in Figure~\ref{fig:test} further show the distributions of the calculated $\Gamma_1$ and $\Gamma_2$ (see Eq.~\ref{eq:gamma}). It is clearly shown that $\Gamma_1$ peaks at the center of the vortex and $\Gamma_2$ has values larger than $2/\pi$ ($\approx0.637$) within the edge of the vortex. Figure~\ref{fig:test}(d) depicts a white-noise velocity field (green arrows) with random values and directions, and the distribution of its corresponding $\Gamma_1$ (yellow-white background). As expected, no vortex has been detected by ASDA in this case. 

\begin{figure*}[tbh!]
\centering
\includegraphics[width=\hsize]{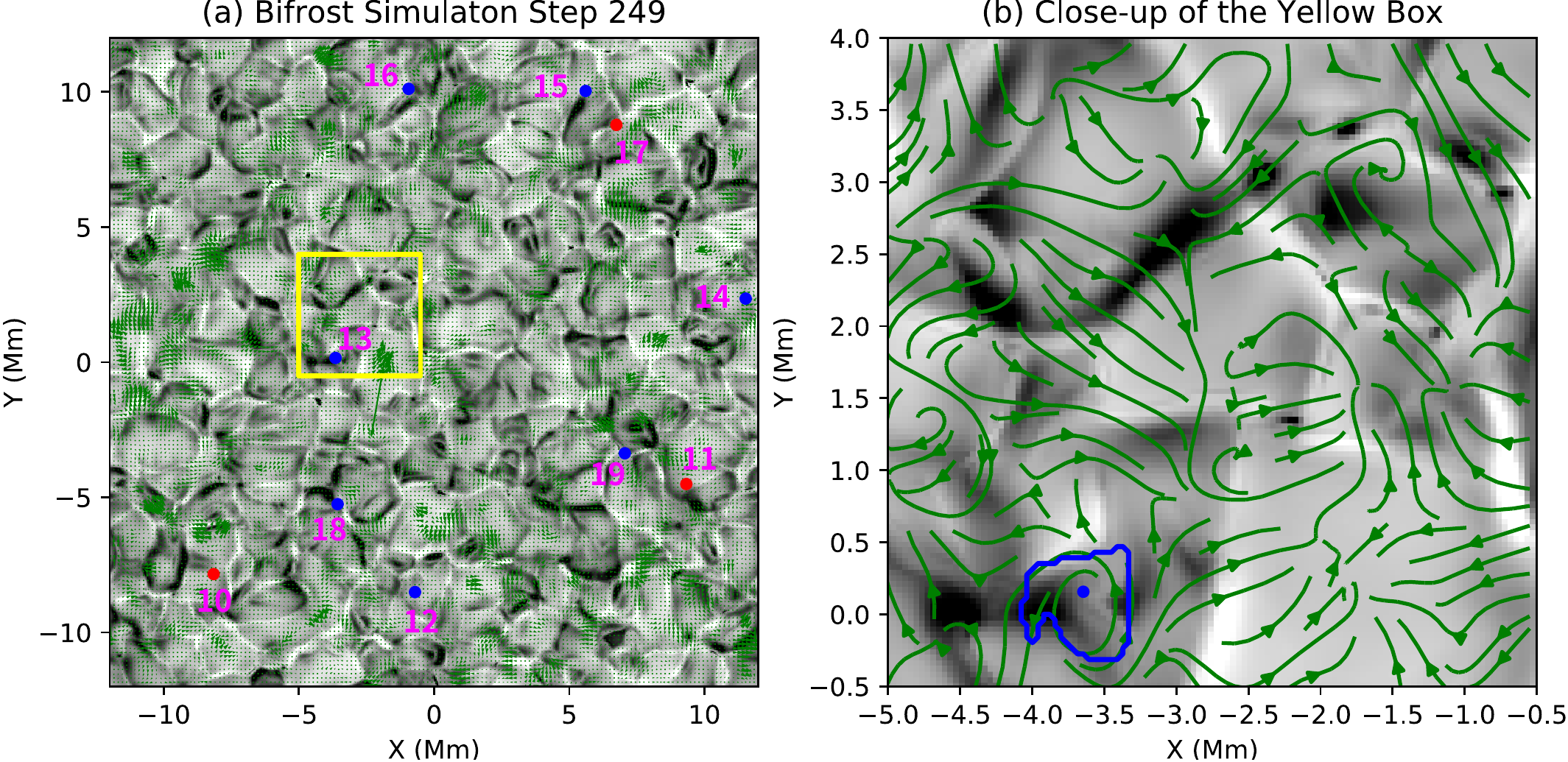}
\caption{Similar to Fig.~\ref{fig:bifrost_full} but with pixel resolution downgraded to the same value (39.2 km) as the SOT observations.}
\label{fig:bifrost_sot}
\end{figure*}

Next, we combine Lamb-Oseen vortices and random white noise to generate a more ``realistic'' synthetic velocity map, more closely comparable to observational data. We generate 1000 Lamb-Oseen vortices with radii $r_{max}$ and rotating velocities $v_{max}$ obeying the following Gaussian distribution:

\begin{equation}
f(x) = \frac{1}{\sqrt{2\pi}\sigma}exp(-\frac{(x-\mu)^2}{2\sigma^2}).
\label{eq:gauss}
\end{equation}

\noindent Here, $f(x)$ is the probability density of the variable $x$, $\mu$ its expected value and $\sigma$ its standard deviation. According to the statistical results of photospheric swirls detailed in Sec. ~\ref{sst}, we set the expected value and standard deviation of the vortices' radii as $\mu_r=7.2$ px and $\sigma_r=1.6$ px, respectively. The expected rotating speed and its standard deviation are set as $\mu_v=0.7$ px per frame and $\sigma_v=0.3$ px per frame, respectively. Next, all the generated vortices are randomly divided into two groups of the same size: one with positive (anti-clockwise) rotation and the other with negative (clockwise) rotation. A background noise map with a size of 5000$\times$5000 px$^2$ is then created with velocities at each pixel having random directions and random magnitudes ranging from 0 to 10\% of $\mu_v$. Finally, all 1000 vortices are randomly inserted into the above background noise map, without any overlap between each vortex. Detection results of vortices from the synthetic data with a noise level of 10\% are shown in Figure~\ref{fig:syn}.

We find at a noise level of 10\%, 96.3\% of the inserted vortices have been successfully detected (Fig.~\ref{fig:syn}), with 100\% accuracy of the rotating direction and 0\% false detection rate. A false detection is when a vortex is detected at a specific location but was not inserted at that location. Figure~\ref{fig:syn}(c) and (d) demonstrate the near perfect recovery of the distribution of the radii and velocities of the inserted vortices by ASDA. Let us now define the location accuracy ($A_l$), radius accuracy ($A_r$) and rotating speed accuracy ($A_s$) of the detection of a swirl located at (x, y) as:

\begin{equation}
\begin{aligned}
& A_l = \frac{|(x, y) - (x_d, y_d)|}{r} \times 100\%, \\
& A_r = \frac{|r_d - r|}{r} \times 100\%, \\
& A_s = \frac{|\mathbf{v}_{rd} - \mathbf{v}_r|}{|\mathbf{v}_r|} \times 100\%,
\end{aligned}
\label{eq:acc}
\end{equation}

\noindent where, $(x_d, y_d)$, $r_d$ and $\mathbf{v}_{rd}$ are the detected location, radius and rotating speed of the swirls, respectively. It turns out that the average location accuracy, radius accuracy and rotating speed accuracy of all vortices detected at a noise level of 10\% are 100\%, 96.7\% and 98.8\%, respectively. Table~\ref{tab:syn} lists the detection rate, false detection rate, average location accuracy, average radius accuracy and average rotating speed accuracy of the all vortices detected at velocity noise levels of 0\%, 10\%, 25\%, 50\%, 75\% and 100\%. It is shown that, the detection rate drops quickly after the noise level reaches 25\%, but still obtains a value of 43\% when the noise is comparable to half of the signal. The false detection rate is consistently zero with increasing noise level, indicating that even though the noise is comparable to the signal, it is still unlikely that ASDA would find a vortex at a location where there is none. Moreover, the detection maintains high accuracies for the location, radius and rotating speed of vortices, even when the velocity noise is comparable to the rotating speed of vortices.

From the results obtained by applying ASDA to the above series of synthetic data, we conclude that our method may detect less swirls than are actually present in data but will return close to zero false detections, even when noise is present in the data.

\section{Application to Numerical Simulation and Observational Data} \label{app}

\begin{figure*}[htb]
\centering
\includegraphics[width=0.8\hsize]{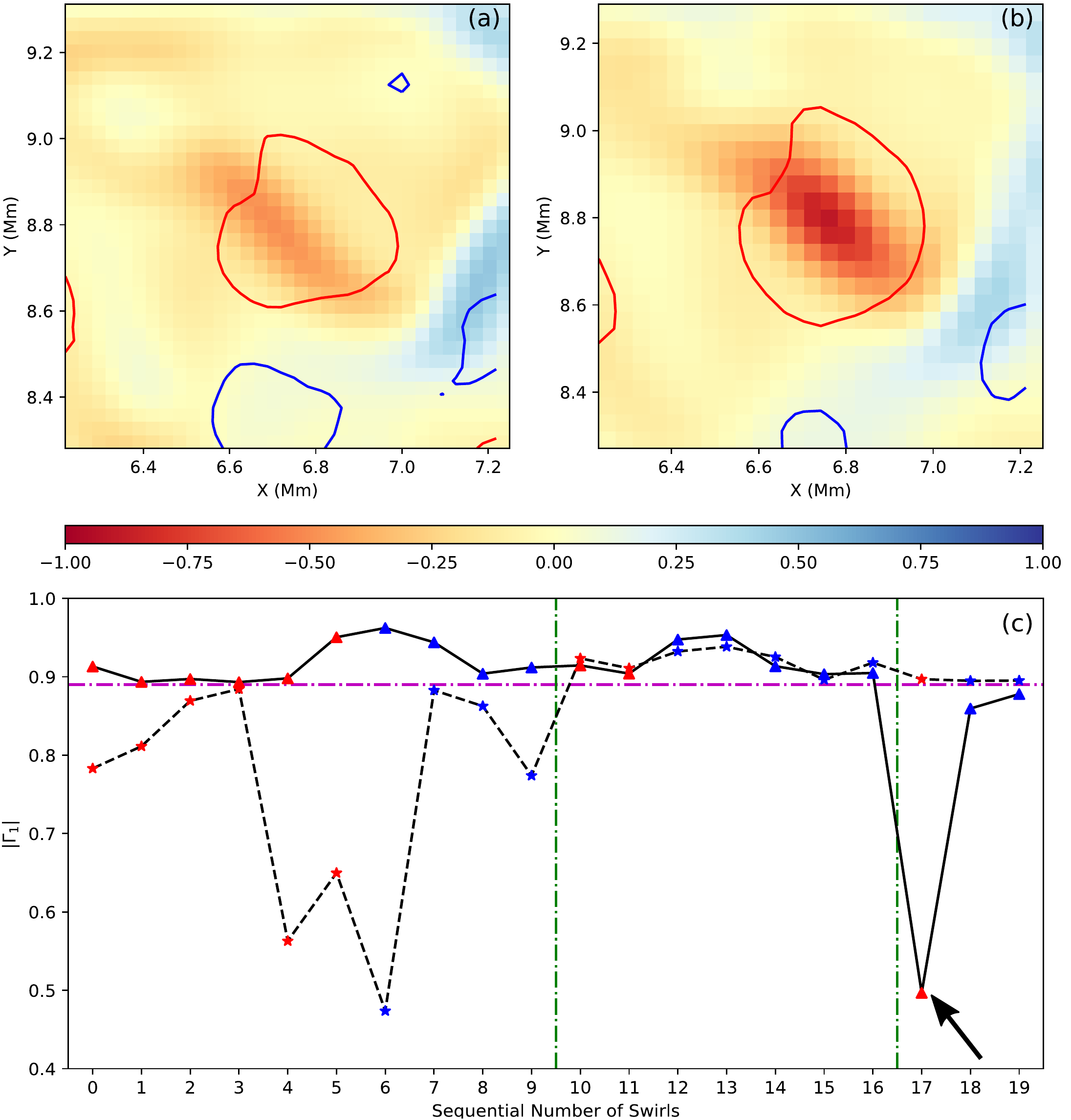}
\caption{\textbf{(a)} and \textbf{(b)} Distributions of $\Gamma_1$ (background)  and $\Gamma_2$ (contours) in the neighborhood of swirl Nr. 17 (Fig.~\ref{fig:bifrost_sot}a), computed from the original (panel a) and downgraded (panel b) Bifrost numerical simulation data, respectively. \textbf{(c)}: Absolute peak $\Gamma_1$ values of all 19 swirls (Fig.~\ref{fig:bifrost_full}a and Fig.~\ref{fig:bifrost_sot}a), with the solid curve and dashed curve from the original and downgraded Bifrost numerical simulation data, respectively. Blue (red) dots are positive (negative) swirls. The horizontal dash-dotted purple line represents a $|\Gamma_1|$ value of 0.89. All swirls left of the left-hand vertical line have higher peak $|\Gamma_1|$ in the original data, and all swirls right of the right-hand vertical line have higher peak $|\Gamma_1|$ in the downgraded data.}
\label{fig:bifrost_compare}
\end{figure*}

\begin{figure*}[htb]
\centering
\includegraphics[width=0.8\hsize]{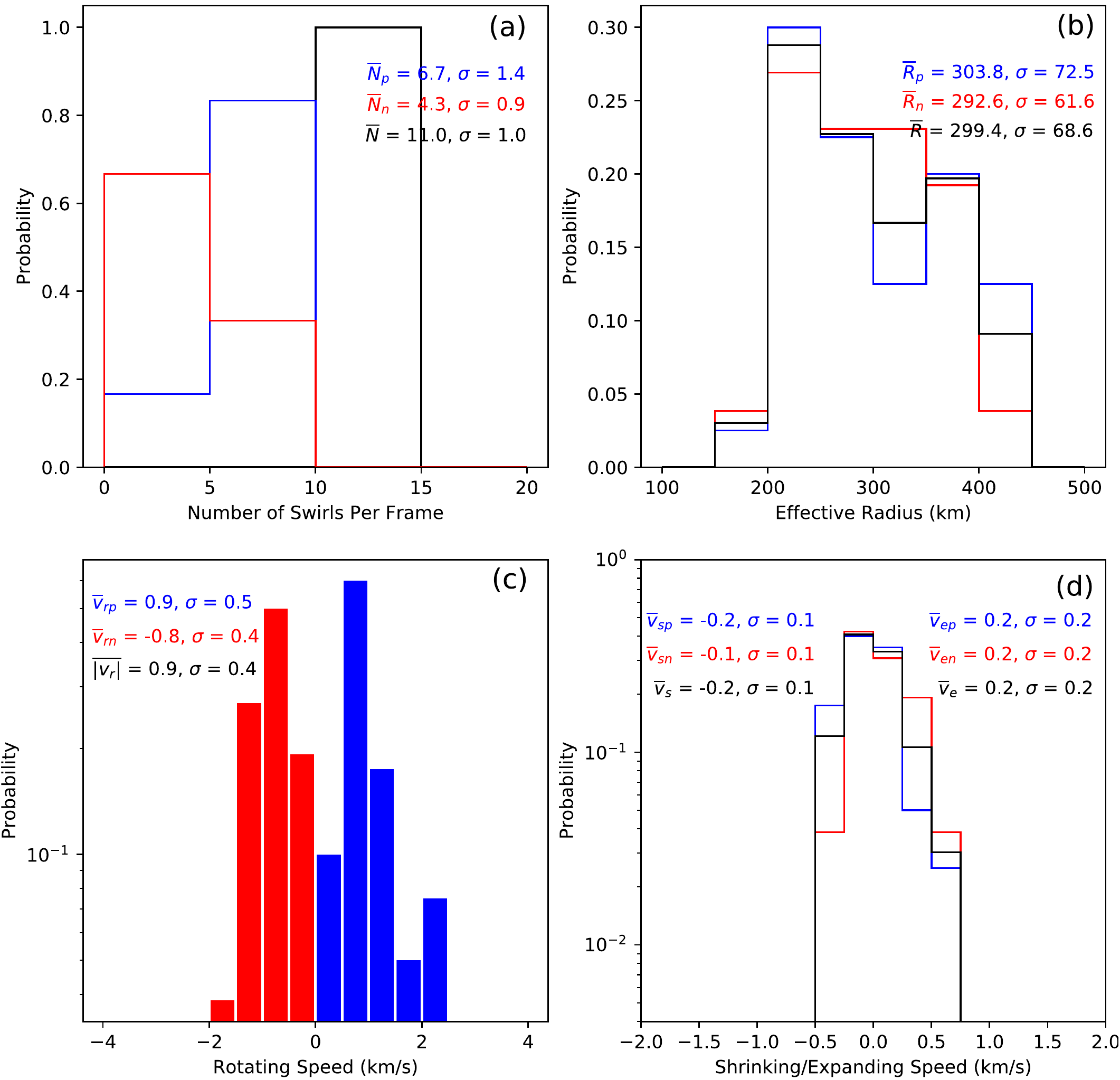}
\caption{Similar to Fig.~\ref{fig:bifrost_stat_full} but for all 64 photospheric swirls detected by ASDA from the downgraded Bifrost numerical simulation data.}
\label{fig:bifrost_stat_sot}
\end{figure*}

\subsection{Bifrost Photospheric Density Swirls} \label{bif}

Before applying ASDA to observational solar data, let us test our method with realistic numerical simulation results obtained with the Bifrost code \citep{Gudiksen11}. Bifrost is a 3D radiation MHD code, which solves the MHD equations on a staggered grid using a 5th/6th order compact finite difference scheme \citep{Carlsson16}. It has been shown to be a very general MHD code, providing a number of modules for boundary conditions and the equation-of-state. More details of the Bifrost simulation can be found in \cite{Gudiksen11}.

The specific simulation we use in this work is publicly available at the {\it Hinode Science Data Centre Europe} website (\url{http://sdc.uio.no/search/simulations}). This simulation provides all variables within the computational domain with a cadence of 10 s. The computational domain has a size of $24\times24\times24$ Mm$^3$ with a 31.25 km horizontal resolution, and $768\times768\times768$ grid points. The average unsigned magnetic field strength in the photosphere (defined as $\tau_{5000}=1$) is set to 40 G. At the beginning of the simulation, a horizontal field of 200 G along the {\it y}-axis is inserted at 2.5 Mm below the photosphere in the convection zone. Later in the simulations, this flux rises to the surface and interacts with the existing field as emerging flux. Thus, we only use results from the first few time steps (249-255) of the simulation, when the inserted large-scale magnetic field had not influenced the photosphere yet. To mimic results obtained by applying ASDA to photospheric intensity ($I$) observations, we use the mass density ($\rho$) in the photosphere from the above simulation as the input of ASDA, considering $I\propto\rho$ for optically thick regions of the solar atmosphere.

Figure~\ref{fig:bifrost_full}(a) shows an example of the photospheric mass density distribution (black-white), overlaid with the estimated velocity field estimated by FLTC (green arrows) and the center locations of detected swirls (blue and red dots) at the simulation time step 249. 9 positive and 8 negative swirls are detected in this panel with a field of view (FOV) of $24\times24$ Mm$^2$. Figure~\ref{fig:bifrost_full}(b) presents a close-up view of the yellow box in panel (a), showing the edges and centers of two swirls with one positive and the other negative. We find that even though these data are not as perfect as the ideal vortices in Figure~\ref{fig:syn}(b), the detected centers and edges match the velocity field well.

Statistical results of 111 swirls detected from 6 velocity field maps generated by FLCT from all 7 density maps (step 249 to step 255) are shown in Figure~\ref{fig:bifrost_stat_full}. Here, $N$, $R$, $v_r$, $v_e$ and $v_s$ represent the number of swirls per frame, effective radius, rotating speed, expanding speed and shrinking speed, respectively. Subscripts $p$ and $n$ stand for positive and negative swirls, respectively. The effective radius $R$ of a swirl is defined as the radius of a circle which has the same area as the swirl :

\begin{equation}
R = \sqrt{\frac{A}{\pi}}.
\label{eq:er}
\end{equation}

\noindent Here, $A$ is the area of the swirl (surrounded by its edge). $v_r$, $v_e$ and $v_s$ are the average values of all points located at the edge of the swirl. We find, in each frame, there are on average 18.5 swirls, with 10.8 positive ones and 7.7 negative ones. The above result suggests a number density of $3.21\times10^{-2}$ Mm$^{-2}$ for photospheric swirls or, in other words, we expect around $1.95\times10^{5}$ swirls at any time in the solar photosphere. 

\begin{figure*}[htb]
\centering
\includegraphics[width=0.8\hsize]{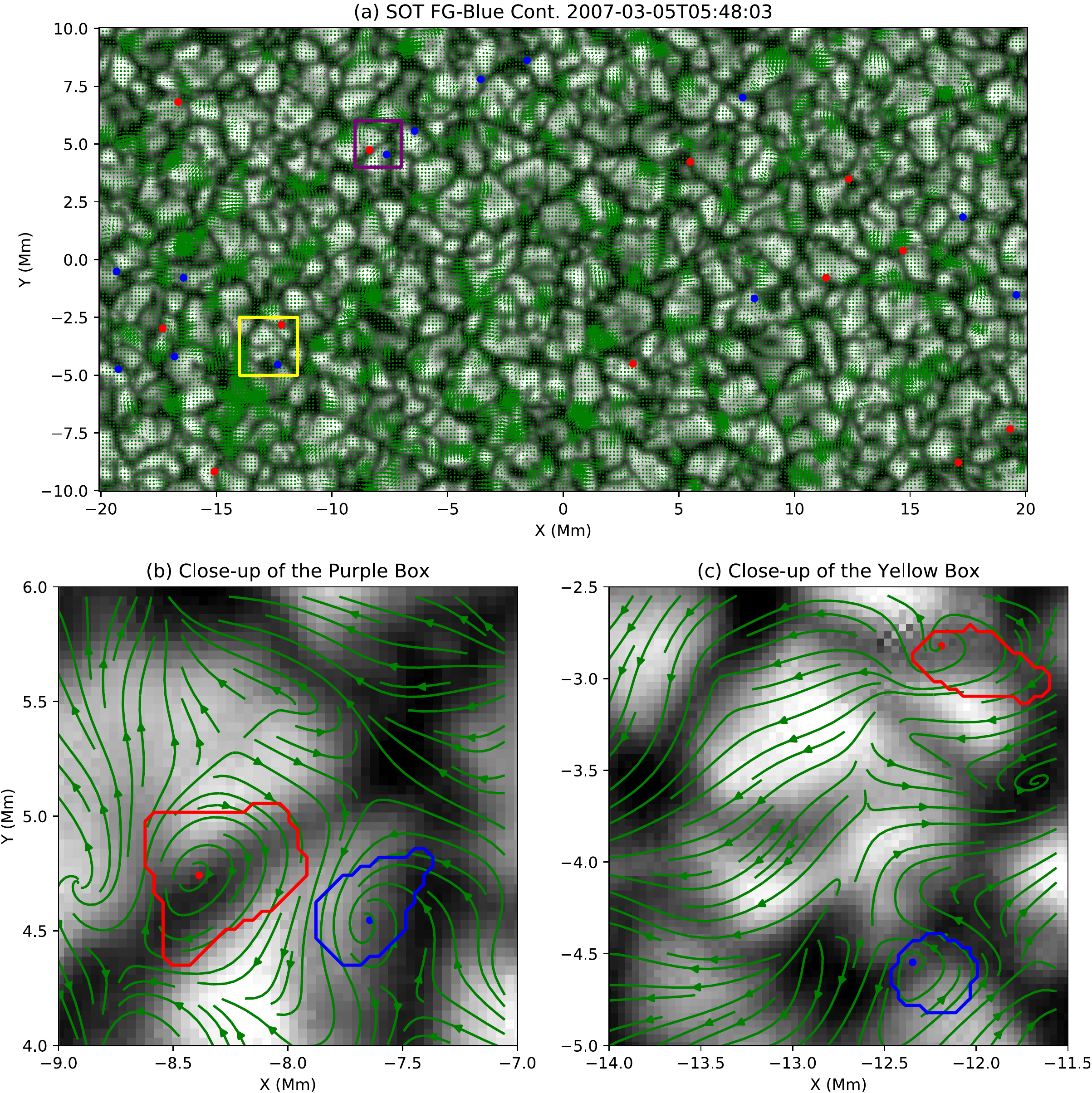}
\caption{Similar to Fig.~\ref{fig:bifrost_full} but for photospheric intensity swirls detected from the SOT FG blue continuum observation at 05:48:03 UT on the 5th March 2007. Panel (b) plots the purple box from panel (a) and panel (c) plots the yellow box from panel (a).}
\label{fig:sot_swirl}
\end{figure*}

Positive and negative swirls show no statistical difference in their effective radius (Fig.~\ref{fig:bifrost_stat_full}b), rotating speed (Fig.~\ref{fig:bifrost_stat_full}c), and shrinking/expanding speed (Fig.~\ref{fig:bifrost_stat_full}d). The average radius and rotating speed are found to be around 240 km and 1.0 km s$^{-1}$. Approximately half of the swirls experience expanding and the other half shrinking. Their shrinking/expanding speeds are small, with absolute values no more than 1 km s$^{-1}$ and average value around 0.2 km s$^{-1}$.

To study the influence of spatial resolution on the detection results of swirls, we downgrade the simulation data to a pixel size of that of the SOT ($\sim$39.2 km, Sect.~\ref{sot}). Let us name the downgraded simulation data as DG1 (downgraded data 1). Figure~\ref{fig:bifrost_sot} is similar to Figure~\ref{fig:bifrost_full}, but shows swirls detected from the downgraded density data. Clearly, comparing to the swirl detection results from the original data, ASDA finds less swirls from the downgraded data with 7 (was 9) positive and 3 (was 8) negative swirls. Out of 17 swirls detected in the original data, 7 are detected (Nr. 10 to 16 in Figs.~\ref{fig:bifrost_full} and ~\ref{fig:bifrost_sot}) in the downgraded data. However, interestingly, there are three more swirls (Nr. 17 to 19) found in the downgraded data, which were not found in the original data. One concern about this is that: is ASDA ``generating'' new swirls when using low-resolution data?

Figure~\ref{fig:bifrost_compare}(a) and (b) show the distributions of $\Gamma_1$ and $\Gamma_2$ in the neighborhood of swirl Nr. 17 (indicated by the black arrow in Fig.~\ref{fig:bifrost_compare}c), with the background $\Gamma_1$ and contours $\Gamma_2$ at levels of $\pm2/\pi$ (blue for positive and red for negative). In both panels, we can find enhanced $\Gamma_1$ within the region surrounded by the red contour (center of the images), which denotes the edge of the swirl. However, the peak value of $\Gamma_1$ is -0.50 in the original data and is enhanced to -0.90 in the downgraded data. The above results suggest that ASDA did not ``make up'' new swirls using the downgraded data, but made some weak swirls strong enough (with absolute value above 0.89) to be identified.

The detection of ``enhanced'' swirl signatures is further investigated in Figure~\ref{fig:bifrost_compare}(c). Red (blue) dots represent negative (positive) swirls. The solid curve shows the absolute peak values of $\Gamma_1$ for all 19 swirls calculated from the original data, and the dashed curve plots the absolute peak values of $\Gamma_1$ from the downgraded data. The horizontal dash-dotted line represents a value of 0.89. It is shown that 52.6\% of the swirls (Nr. 0 to 9) are weakened below the threshold after downgrading the spatial resolution. 36.6\% (Nr. 10 to 16) keep similar peak $\Gamma_1$ values and thus are detected in both the original and downgraded data. 15.8\% (Nr. 17 to 19) are enhanced from below to above the threshold.

\begin{figure*}[htb]
\centering
\includegraphics[width=0.8\hsize]{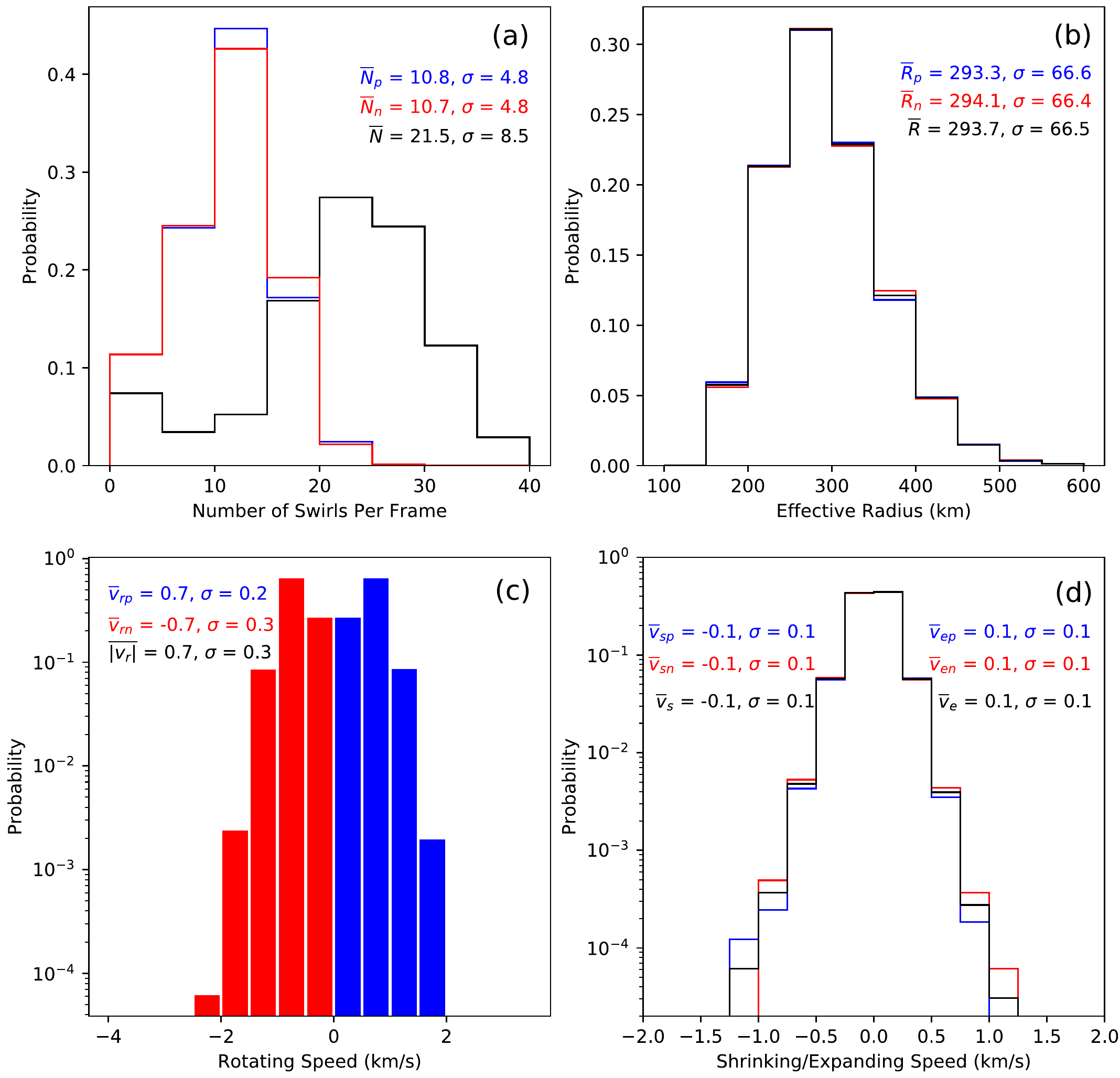}
\caption{Similar to Fig.~\ref{fig:bifrost_stat_full} but for all 32590 photospheric intensity swirls detected from the SOT FG blue continuum observations on the 5th March 2007.}
\label{fig:sot_stat}
\end{figure*}

In total, ASDA detects 66 swirls (was 111) from all 6 velocity fields estimated based on the downgraded data. A similar statistical study has also been performed as was conducted on the original data, and is shown in Figure~\ref{fig:bifrost_stat_sot}. On average, there are 11.0 swirls in each frame (was 18.5) of the downgraded data, indicating a number density of $\sim1.91\times10^{-2}$ Mm$^{-2}$ or a total number of $1.16\times10^5$ swirls in the photosphere. The average radius is enlarged from 240 km to $\sim300$ km. The detected average rotating speed and expanding/shrinking speed remain almost unchanged.

We then further downgrade the data to the same resolution as data from the SST/CRISP instrument ($\sim43.6$ km, Sect.~\ref{sst}) and name these downgraded data DG2 (downgraded data 2). On average, there are 8.8 swirls in each frame detected from the further downgraded data, indicating a number density of $\sim1.53\times10^{-2}$ Mm$^{-2}$ or a total number of $0.93\times10^5$ swirls in the photosphere. The average radius is further enlarged to $\sim330$ km. The detected average rotating speed and expanding/shrinking speed still remain almost unchanged.

The above comparisons between the original and two downgraded numerical simulation datasets suggest a vital influence of the spatial resolution on the number and radii of swirls detected. It is, therefore, likely that current observational data will only allow us to calculate upper estimates for the radii and lower estimate for the total number of swirls in the solar atmosphere.

\subsection{SOT Photospheric Intensity Swirls} \label{sot}

The {\it Solar Optical Telescope} \citep[SOT,][]{Tsuneta08} onboard the Hinode \citep{Kosugi07} satellite was launched in 2006. SOT consists of two main parts: the main 50 cm aperture telescope (Optical Telescope Assmebly, OTA) and the Focal Plane Package (FPP). The SOT FPP contains three CCD cameras including the Filtergraph (FG), Spectro-polarimeter (SP) and Correlation Tracker (CT). The Broadband Filter Imager (BFI) of the OTA produces photometric images in 6 bands in a wavelength range from 388 nm to 668 nm with a spatial resolution as high as $\sim0.054$\arcsecs and a rapid cadence of less than 10 s.

\begin{figure*}[htb]
\centering
\includegraphics[width=\hsize]{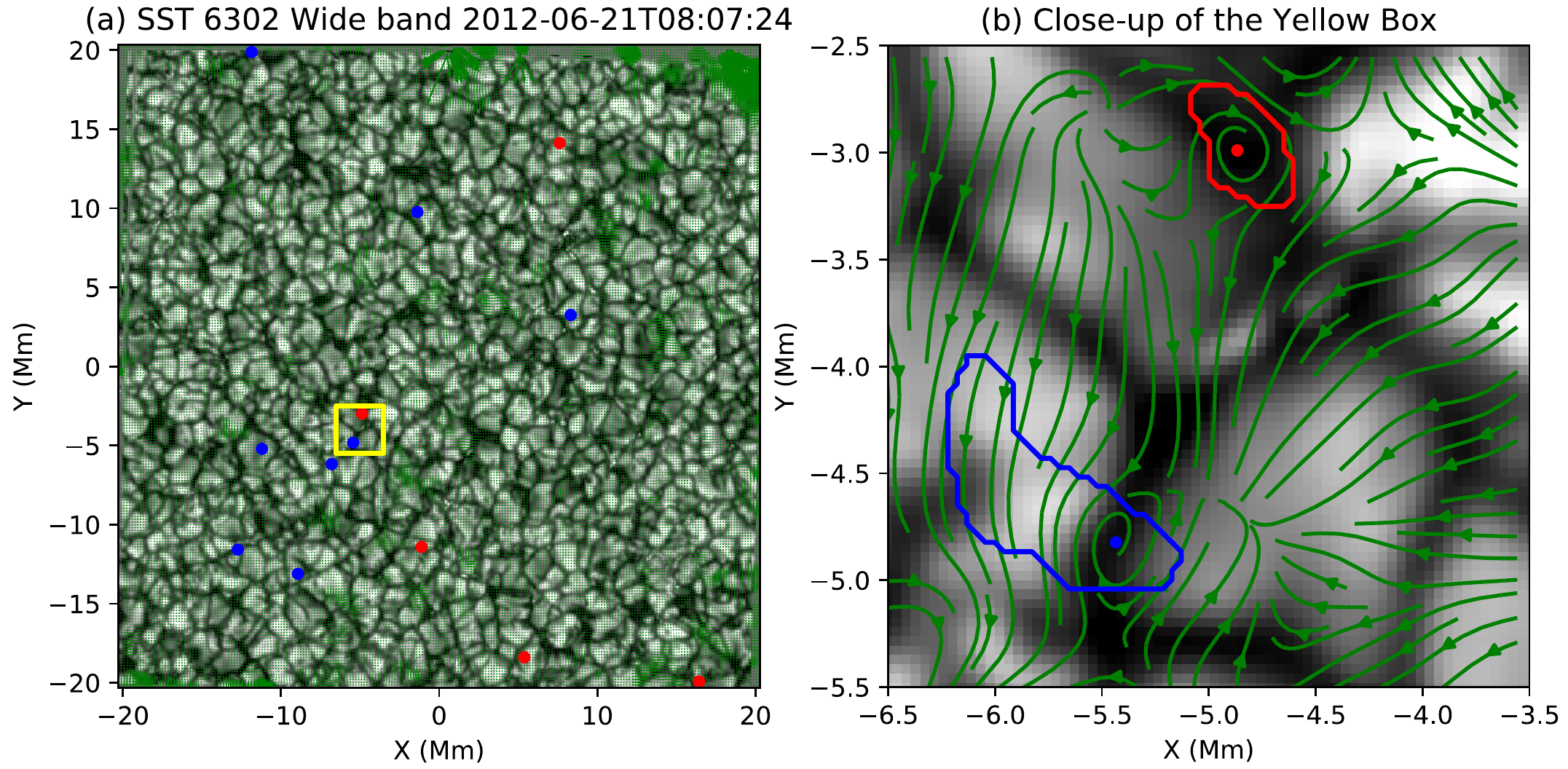}
\caption{Similar to Fig.~\ref{fig:bifrost_full} but for photospheric intensity swirls detected from the SST \ion{Fe}{1} 6302 \AA\ wide-band observation at 08:07:24 UT on the 21st June 2012.}
\label{fig:sst_swirl}
\end{figure*}

The observational data to be studied in this subsection consists of blue continuum images sampled with the wide-band imager centered at a wavelength of 450.4 nm with a band width of 0.4 nm. These data were collected between 05:48:03 UT and 08:29:59 UT on the 5th March 2007. 1515 images were taken during the above period with a cadence of $\sim6.4$ s, and a FOV centered at $x_c$=5.3\arcsecs, $y_c$=4.1\arcsecs. Each of the images has a size of 1024$\times$512 px$^2$ with a pixel size of 0.0545$\arcsecs$ (39.2 km) and, hence, a FOV of $\sim56\arcsecs\times28\arcsecs$ (40.1 Mm $\times$ 20.1 Mm). The level-1 fits files were then processed with the ``fg\_prep.pro'' program provided in the {\it SolarSoft IDL} packages, to correct camera readout defects, and to apply the dark-current and flat-field corrections.

An example of the blue continuum photospheric observations, the velocity field estimated by FLCT and the swirls detected by ASDA at 05:48:03 UT is shown in Figure~\ref{fig:sot_swirl}(a). 13 positive (blue) and 12 negative (red) swirls have been detected, with Figure~\ref{fig:sot_swirl}(b) and (c) showing close-up views of four swirls. It is clear that most of the swirls are located within intergranular lanes, as expected from previous studies \citep[e.g.,][]{Bonet08, Shelyag11}.

Statistical results of the 32590 swirls detected from 1514 FLCT velocity field maps generated from all 1515 SOT FG blue continuum photospheric observations are shown in Figure~\ref{fig:sot_stat}. Here, again, $N$, $R$, $v_r$, $v_e$ and $v_s$ represent the number of swirls per frame, effective radius, rotating speed, expanding speed and shrinking speed, respectively. Subscripts $p$ and $n$ stand for positive and negative swirls, respectively. On average, there are 21.5 swirls detected in each frame, with 10.8 positive ones and 10.7 negative ones. The detection results show no preference for the number of positive and negative swirls in the solar photosphere. The above results indicate a number density of $2.67\times10^{-2}$ Mm$^{-2}$ or a total number of $1.62\times10^{5}$ swirls in the solar photosphere. This indicates that we have detected $\sim40\%$ more swirls from the SOT observations than the downgraded simulation data DG1 with the same resolution, implying that the Bifrost simulation photospheric density data might have underestimated the number of swirls in the photosphere.

The average effective radius and average rotating speed of all swirls detected from the SOT FG blue continuum observations are $\sim293$ km and $\sim0.7$ km s$^{-1}$, without any positive or negative preference, which are comparable to the values obtained using the downgraded numerical simulation data DG1. Most of the swirls experience no significant expanding/shrinking motions, with average expanding/shrinking speeds of $\sim0.1$ km s$^{-1}$.

For a detected swirl, we use the following simple principle to determine whether it is located in intergranular lanes or not: if the average intensity of all pixels within the swirl is smaller than the average intensity of all the SOT photospheric data obtained, it is believed to be located in intergranular lanes. It is then found that 71.5\% of the positive and 71.2\% of the negative swirls are located in intergranular lanes. We should note that, because swirls are spatially extended and the photospheric intensity are not uniform, the number of intergranular-lane swirls have been most likely underestimated.

\subsection{SST Photospheric Intensity Swirls} \label{sst}

\begin{figure*}[htb]
\centering
\includegraphics[width=0.8\hsize]{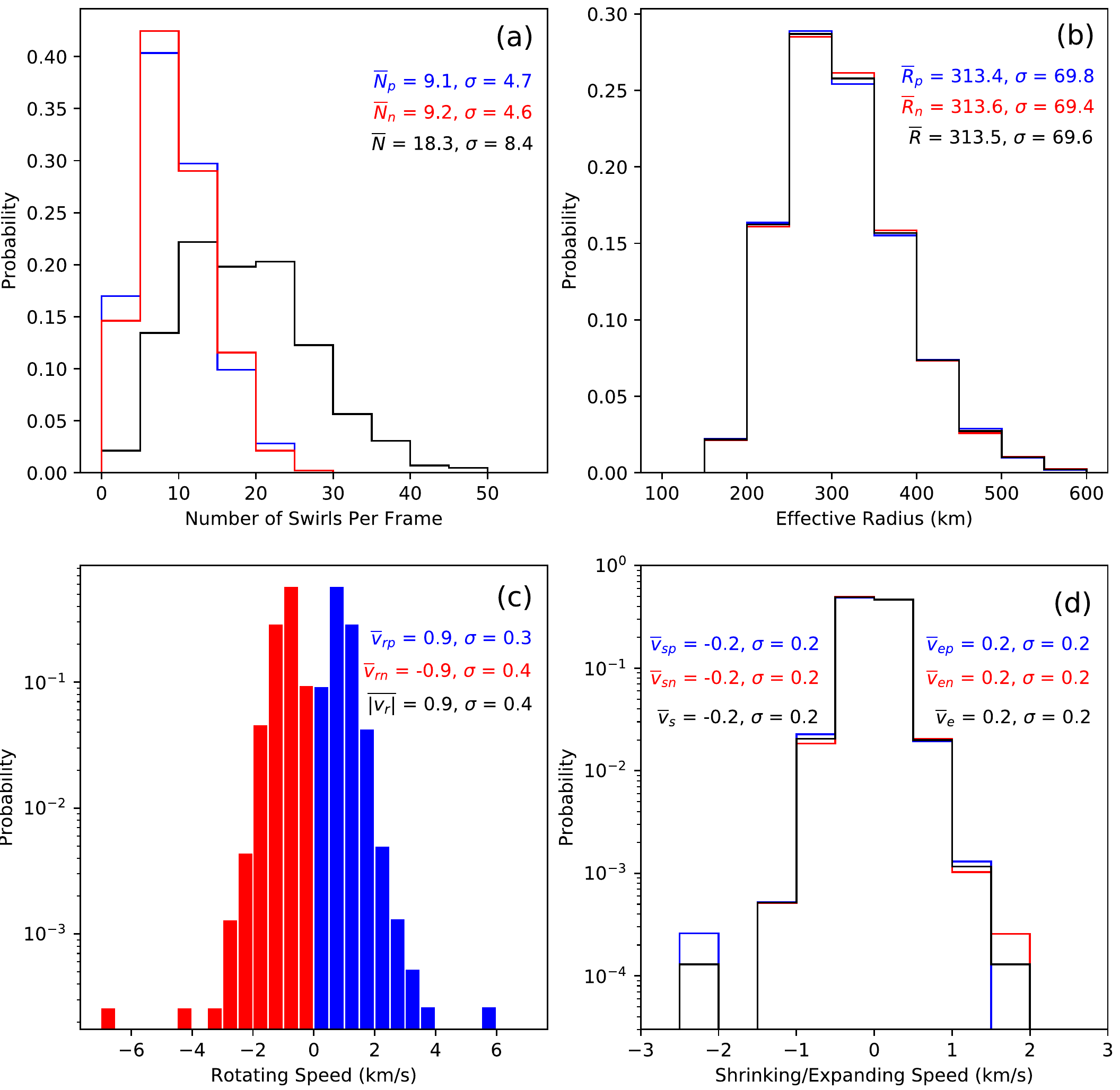}
\caption{Similar to Fig.~\ref{fig:bifrost_stat_full} but for all 7741 photospheric intensity swirls detected from the SST \ion{Fe}{1} 6302 \AA\ wide-band observations on the 21st June 2012.}
\label{fig:sst_stat}
\end{figure*}

The observational data analysed in this sub-section were sampled by the SST/CRISP instrument on the $21$st June 2012 between $08$:$07$:$22$ UT and $09$:$05$:$44$ UT. The SST/CRISP sequence which ran during this time consisted of an eleven-point H$\alpha$ line scan (sampling evenly between $\pm 1.3$ \AA\ from the line core), a nineteen-point \ion{Ca}{2} $8542$ \AA\ line scan (sampling evenly between $\pm 0.5$ \AA\ from the line core), and a single-point full-Stokes measurement at the core of the \ion{Fe}{1} $6302.5$ \AA\ line. The $55$\arcsecs$\times55\arcsecs$ (40.6 Mm $\times$ 40.6 Mm) FOV sampled the quiet Sun close to the disk center, with initial central co-ordinates of $x_\mathrm{c}$=$-3$\arcsecs, $y_\mathrm{c}$=$70$\arcsecs. Wide-band images were collected co-temporally for each wavelength.

The data were reduced using the Multi-Object Multi-Frame Blind Deconvolution \citep[MOMFBD, ][]{Noort05} method and the standard CRISPRED pipeline (\citealt{Rodriguez15}), including additional steps to account for differential stretching \citep[suggested by][]{Henriques12}. After applying these reductions, the data had a science-ready cadence of approximately $8.25$ s and a pixel scale of $0.059\arcsecs$ (43.6 km), matching the properties of data required to observe swirls as proposed by \cite{Kato17}. To make direct comparison with the results obtained from the (downgraded) numerical simulations and SOT observational data, we use the wide-band images of the \ion{Fe}{1} $6302.5$ \AA\ line to demonstrate photospheric swirl detection.

An example of the \ion{Fe}{1} $6302.5$ \AA\ wide-band images, FLCT velocity field and swirls detected by ASDA at 08:07:24 UT is shown in Figure~\ref{fig:sst_swirl}(a). 8 positive (blue) and 5 negative (red) swirls have been detected, with Figure~\ref{fig:sst_swirl}(b) depicting a close-up view of two swirls. Again, similar to the SOT photospheric swirls, it turns out that most of the SST photospheric swirls are located in intergranular lanes.

There are in total 7741 swirls detected from all 424 velocity maps derived from 425 \ion{Fe}{1} $6302.5$ \AA\ wide-band images, among which 3840 (49.6\%)/3901 (50.4\%) ones have positive/negative rotations. Using the method described in the second but last paragraph in Sect.~\ref{sot}, we find that at least 66.5\% of positive and 69.4\% of negative photospheric swirls are located in intergranular lanes - supporting the result found from the SOT photospheric swirls.

\begin{table*}[ht]
\centering
\begin{tabular}{|c|c|c|c|c|c|c|}
\hline
Data& Pixel Size & $\overline{R}$ & $\overline{|v_r|}$  & $\overline{|v_{e}|}$ & Number Density & Total Number in\\
& km & km & km s$^{-1}$ & km s$^{-1}$ & $10^{-2}$ Mm$^{-2}$ & Photosphere ($10^5$)\\
\hline
Bifrost & 31.25 & 239.5 & 1.0 & 0.2 & 3.21 & 1.95 \\
Bifrost & 39.2 & 299.4 & 0.9 & 0.2 & 1.91 & 1.16 \\
Bifrost & 43.6 & 330.2 & 0.9 & 0.2 & 1.53 & 0.93 \\
{\it Hinode}/SOT & 39.2 & 293.7 & 0.7 & 0.1 & 2.67 & 1.62 \\
SST/CRISP & 43.6 & 313.5 & 0.9 & 0.2 & 1.11 & 0.68 \\
\hline
\end{tabular}
\caption{\label{tab:sum} Average effective radius ($\overline{R}$, Eq.~\ref{eq:er}), average rotating speed ($\overline{|v_r|}$), average shrinking/expanding speed ($\overline{|v_{e}|}$), number density and expected total number in the photosphere at any moment of time of swirls detected in different data employed in this article.}
\end{table*}

On average, there are 18.3 swirls detected in each frame, with 9.1 positive and 9.2 negative ones. This indicates a number density of $1.11\times10^{-2}$ Mm$^{-2}$ or a total number $0.68\times10^{5}$ of swirls in the solar photosphere, which is about 70\% of what we obtained from the spatial downgraded Bifrost simulation data with the same resolution (DG2). The average effective radius of swirls are slightly higher than SOT photospheric swirls, which could be expected from the lower spatial resolution of the SST/CRISP obserations. The rotating speed and shrinking/expanding speed show similar distribution to SOT photospheric swirls. Again, the SST swirls display no positive or negative preference of rotation.

Statistics of swirls detected in the original Bifrost numerical simulation, two downgraded simulations, SOT and SST observations are summarized in Table ~\ref{tab:sum}.

\subsection{Lifetime of Swirls} \label{life}

\begin{figure*}[htb]
\centering
\includegraphics[width=\hsize]{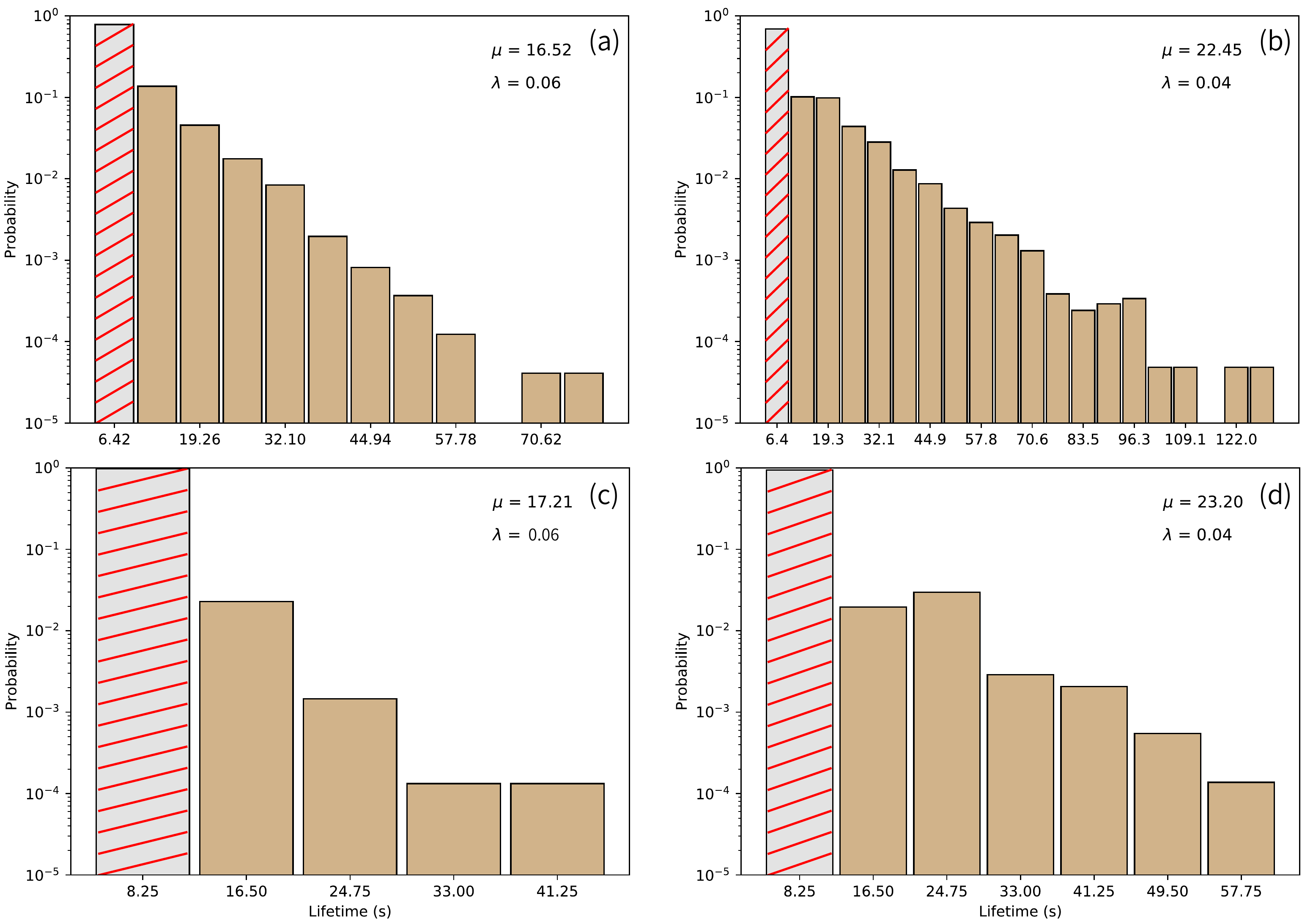}
\caption{\textbf{(a)} and \textbf{(c)} Distributions of lifetime of SOT and SST photospheric swirls. \textbf{(b)} and \textbf{(d)} Similar to (a) and (c), but allowing one frame missing when estimating the lifetimes (see Sect.~\ref{life}). Gray bars with red slashes represent swirls that appear in only one frame. {$\mu$ and $\lambda$ in each panel are the average lifetime and the MLE of the exponential rate parameter (see Eq.~\ref{eq:mle}), respectively.}}
\label{fig:lifetime}
\end{figure*}

Suppose there are two swirls (S$_1$ and S$_2$) to be studied. They are detected in two successive frames, i.e., S$_1$ is detected at time $t_0$ and S$_2$ at time $t_0+\Delta t$, where $\Delta t$ is the cadence of the observation. S$_1$ and S$_2$ are then considered as the same swirl, if the expected location of the center of S$_1$ after $\Delta t$ is located within S$_2$:

\begin{equation}
\mathbf{c_1} + \mathbf{v_{c1}}\cdot \Delta t \subset S_2.
\label{eq:lif}
\end{equation}

\noindent Here, $\mathbf{c_1}$ is the location of the center of swirl S$_1$, and $v_{c1}$ the speed of the center of $S_1$. By applying the above algorithms to all swirls detected, we are then able to determine their lifetimes. Figure~\ref{fig:lifetime}(a) and (c) show the distributions of lifetimes of all SOT and SST photospheric swirls. We shall notice that the lifetime estimation of swirls which appear in only one frame (represented by gray bars with red slashed in Fig.~\ref{fig:lifetime}) are not fully reliable. We omit them in the following estimations. Even though we can find few SOT photospheric swirls (with possibilities of the order of $10^{-5}$ to $10^{-4}$) having lifetimes longer than 40 s, there is no significant difference between the lifetimes of swirls detected in SOT and SST. The average lifetimes are 16.5 s and 17.2 s, respectively. 

Considering that a swirl may experience changes to its rotational motion through time, we re-evaluate the lifetimes with the same method described above but allowing swirls to be missing from one frame. Corresponding results of this re-evaluation of SOT photospheric swirls and SST photospheric swirls are shown in Figure~\ref{fig:lifetime}(b) and (d), respectively. Again, the number of both SOT and SST photospheric swirls decreases exponentially with increasing lifetime. The overall distributions show slightly longer lifetimes, with average lifetimes of 22.4 s and 23.2 s, respectively.

It is observed that the logarithm of probability decreases quasi-linearly with increasing lifetime in all the panels in Figure~\ref{fig:lifetime}, no matter which of the above two methods is used. A natural consideration would be performing exponential fits, which, however, could lead to biased and perhaps less accurate results, because they are least-square fitting to the binned data. Thus, we perform the maximum likelihood estimation (MLE), which has been fully studied for power-law fits, is well established for asymmetrical distributions, and also suitable for distributions with exponential characteristics \citep[e.g.][]{Goldstein2004, Newman2005, Bauke2007}:

\begin{equation}
\begin{aligned}
& P(x) \propto e^{-\lambda x}, \\
& \lambda = {\mu}^{-1},
\end{aligned}
\label{eq:mle}
\end{equation}

\noindent where, $x$ is the independent variable (the lifetime of swirls here), of which the possibility ($P$) obeys an exponential distribution. $\lambda$ ($>0$) is the exponential rate parameter, of which the MLE is proved to be the reciprocal of the average value ($\mu$) of $x$ \citep[e.g.][]{Ross2004}. The exponential rate parameters of both SOT and SST photospheric swirls are found to be almost the same (Figure~\ref{fig:lifetime}).

\color{black}

It is intriguing that 94\%-98\% of the swirls have lifetimes less than 16.5 s (gray bars in Fig.~\ref{fig:lifetime}c and d), and 69\%-78\% of the swirls have lifetimes less than 12.84 s (gray bars in Fig.~\ref{fig:lifetime}a and b). This means the majority of detected swirls do not appear in successive frames with current cadence. It is clear that higher resolution data will be required to accurately identify the average lifetimes of swirls. {We note that, we have excluded the leftmost bins when estimating the average lifetimes ($\mu$) and MLE of the exponential rate parameters ($\lambda$). Values of $\mu$ and $\gamma$ are expected to be modified with higher resolution data employed in the future.}

\section{Discussions and Conclusions} \label{conc}

In this article, we have presented a new open-source Automated Swirl Detection Algorithm (ASDA), including tests of it on a single Lamb-Oseen vortex and on a series of synthetic data with 1000 vortices with different levels of noise. ASDA's application is demonstrated by applying it to detect photospheric swirls from realistic numerical simulations, space-borne observations and ground-based observations. We make the source codes of ASDA and the synthetic data publicly available at \url{https://github.com/PyDL/ASDA}. Codes related to the analysis of the simulation and observational data are also available up on request.

To examine the reliability of ASDA, we generated a series of synthetic data containing 1000 Lamb-Oseen vortices with radii and rotating speeds following Gaussian distributions. These vortices are distributed at random positions in a 5000$\times$5000 px$^2$ image containing randomly oriented velocity noises at levels from 0\% to 100\% of the average rotating velocity of the generated vortices. ASDA has shown very high location, radius and speed detection accuracies (mostly above 90\%) and 0 false detection at all noise levels. The detection rate drops from 100\% to 43.1\% and 3.2\% when the noise level increases from 0\% to 50\% and 100\%. The above results indicate that ASDA could miss some swirls if considerable noise is present in the data, but still obtains 0 false detection and very high accuracy in determining the inferred properties of detected swirls.

Applying ASDA to Bifrost simulation data has shown that, with a spatial resolution of 31.25 km, we could expect $1.95\times10^5$ swirls in the solar photosphere at any moment of time, with an average radius of about 240 km. The total number of swirls expected in the photosphere is reduced to $1.16\times10^5$ when the spatial resolution of the simulation is downgraded to 39.2 km (DG1, the same pixel scaling as that of the SOT observations). The average radius is also enlarged to $\sim 300$ km. The total number (average radius) of swirls is further reduced (enlarged) to $0.93\times10^5$ ($\sim 330$ km) when the spatial resolution is further downgraded to 43.6 km (DG2, the same as that of the SST observations). The above results suggest a vital influence of the spatial resolution on the number and radius of swirls detected. We suggest future study on solar atmospheric swirls should use high-resolution observations \citep[e.g., the Visible Broadband Imager onboard the Daniel K. Inouye Solar Telescope will have a spatial resolution as high as 0.022\arcsecs/15 km and be available soon, ][]{Ferayorni16} and keep in mind the influences of the spatial resolution.

The total number of swirls expected in the photosphere at any moment of time and their average radius are found to be $1.62\times10^5$ and $\sim 294$ km, respectively, after applying ASDA to a series of SOT observations. Comparing the above values with those derived from the downgraded numerical simulation data DG1, we conclude that applying ASDA to the Bifrost simulation has underestimated the number of swirls in the photosphere. The SST observations analysed here have revealed only about 70\% of the swirls found in the similar-resolution downgraded numerical simulation data DG2. Namely, applying ASDA to the SST \ion{Fe}{1} 6302 \AA\ wide-band observations has resulted in a total number of $0.68\times10^5$ swirls in the solar photosphere at any moment of time. It should be remembered that one of the main draw-backs of ground-based data, such as that sampled by the SST/CRISP instrument, is variable atmospheric seeing. With regards to ASDA, this means that any random modulations to the incoming wave-front which are not corrected by the adaptive optics or data processing will introduce spurious velocities in the FLCT step. 

We find that, swirls detected by ASDA in data from the numerical simulations, its two downgrades to observational resolutions, as well as SOT and SST observations, show many similarities, despite the different resolutions and seeing effects. Among all the photospheric swirls, half rotate in a counter-clockwise (positive) direction and half rotate in a clockwise (negative) rotation direction. Around 70\% of the swirls are found to be located in intergranular lanes, no matter in which direction they rotate. The rotational speeds of swirls detected from different data are found to be similar, with actual values mostly less than 2 km s$^{-1}$ (3$\sigma$) and average values less than 1 km s$^{-1}$. Most of the swirls experience no significant expanding/shrinking motion. Additionally our results show that, with a similar resolution, one could expect less swirls to be detected by ground-based observations. However, obtaining information on swirl properties (e.g., radius, rotating speed and shrinking/expanding speed) seems to be rarely affected, whichever instrument is used.

We, then, employed two approaches to estimate the lifetime of SOT and SST photospheric swirls. One approach was to investigate successive frames and another allowed one frame missing. SST photospheric swirls have almost the same average lifetime as SOT photospheric swirls, with $\sim 17$ s and $\sim 23$ s from the above two approaches, respectively. Additionally, the number of either SST or SOT swirls shows an exponential decrease with increasing lifetime, no matter which of the two approaches we use. Interestingly, the majority of detected photospheric swirls do not appear in successive frames. 69\%-78\% of SOT and 94\%-98\% of SST photospheric swirls have lifetimes less than twice of their own cadences (12.84 s and 16.5 s, respectively). This suggests that it is likely possible that our estimation of the lifetime of most swirls (and their average lifetime) has been largely affected by the relatively low cadences of the observations. We recommend that further work studying the lifetime of photopsheric swirls should use observations with much higher cadences than currently employed in this paper (6 s), preferably from a space-borne instrument.

ASDA has been fully tested with synthetic, numerical simulation and observational data in this work. The tool has been proved reliable in detecting photospheric swirls from both simulation and observational data. Considering that it is a versatile code which can be utilized in various regimes of solar physics, we list some potentially promising future research directions where ASDA may provide useful insight:

\noindent {\textbf{1.} How are swirls detected by ASDA distributed at different locations? For example, (1) all swirls detected in this paper were located close to the disk center. How will the statistics of swirls be affected if they are located near limb? (2) Given similar observations using the same instrument, do closed-field quiet regions (QRs) have the same swirl density as open-field coronal hole (CH) regions? Do QR and CH both show no preference of rotating direction in either hemisphere?}

\noindent {\textbf{2.} Both observations utilized in this paper were centered close to the disk center and we found no direction preference in swirl directions. Will there be any direction preference if the observation is located further north or south, and will the direction preference change over different phases of a solar cycle, resembling or differing the heliospheric helicity rule \citep[e.g.,][]{Pevtsov1995, Tiwari2009, Miesch2016}? Investigating the above questions will help in having an understanding on the solar dynamo and its relationship with the solar atmosphere.}

\noindent \textbf{3.} Are all (or most of) the solar atmospheric swirls ``tornadoes'' \citep[e.g.,][]{Wedemeyer12}? We propose to apply ASDA to simultaneous photospheric, chromospheric, transition-region and coronal observations, and investigate the overlaps between detected swirls at different heights.

\noindent \textbf{4.} What is the relationship between solar intensity swirls and magnetic swirls \citep[e.g.][]{Shelyag11}? Applying ASDA to simultaneous high-resolution photospheric intensity and magnetic field observations will help in answering this question.

\noindent \textbf{5.} What roles do rotational motions play in driving Rapid Blueshifted/Redshifted Excursions (RBEs/RREs) \citep[e.g.,][]{Kuridze15} and spicules \citep[e.g.,][]{Pontieu07}? Developing an automated detection method of RBEs/RREs from spectral observations, and then carrying out a comparative analysis of the detection results of swirls found by ASDA from simultaneous intensity observations, could contribute to answering this question. This would be a major progress, given RBEs/RREs and spicules may play an important role in the energisation of the lower solar atmosphere, or providing mass flux to solar wind. 

\noindent {\textbf{6.} In \cite{Zank2018}, a coronal heating model was set up with transverse photospheric convective fluid motions driving predominantly quasi-2D (nonpropagating) turbulence in the mixed-polarity ``magnetic carpet'', together with a minority slab (Alfv{\'e}n) component. The idea of energy transport through swirls from the photosphere into the corona is consistent with the above model, and should be further explored.}

\acknowledgments
The authors thank the Science and Technology Facilities Council (STFC, grant numbers ST/M000826/1, ST/L006316/1) for the support to conduct this research. Bifrost is a versatile and flexible 3D Radiation MHD code developed in Oslo. Publicly available simulation results can be downloaded from \url{http://sdc.uio.no/search/simulations}. Hinode is a Japanese mission developed and launched by ISAS/JAXA, with NAOJ as domestic partner and NASA and STFC (UK) as international partners. Hinode is operated by these agencies in co-operation with ESA and NSC (Norway). The Swedish $1$-m Solar Telescope is operated on the island of La Palma (Spain) by the Institute for Solar Physics of Stockholm University in the Spanish Observatorio del Roque de los Muchachos of the Instituto de Astrof\'isica de Canarias.

\bibliographystyle{yahapj}
\bibliography{references}

\end{document}